\documentclass[aps,prx,twocolumn,superscriptaddress]{revtex4-2}
\usepackage{graphicx,amssymb,amsmath,amsfonts,empheq}
\usepackage[dvipsnames]{xcolor}
\usepackage{color}
\usepackage{braket}
\usepackage{latexsym}
\usepackage{upgreek}
\usepackage{dsfont}
\usepackage{bm}
\usepackage{multirow}
\usepackage{enumitem}
\usepackage[normalem]{ulem}
\usepackage{url}
\usepackage{hyperref}
\hypersetup{
colorlinks = true,
linkcolor = [rgb]{0,0,0}, 
citecolor = [rgb]{0.13,0.55,0.13},
urlcolor = [rgb]{0.25, 0.41, 0.88}}
\usepackage{bbm}

\usepackage{tikz}
\usepackage{tikz-cd}
\usetikzlibrary{arrows}
\usetikzlibrary{intersections}
\usetikzlibrary{shapes.geometric}
\usetikzlibrary{decorations.pathmorphing, patterns,shapes}
\usetikzlibrary{decorations.markings}

\tikzset{
	partial ellipse/.style args={#1:#2:#3}{
		insert path={+ (#1:#3) arc (#1:#2:#3)}
	}
}

\tikzset{
	mid arrow/.style={postaction={decorate,decoration={
				markings,
				mark=at position .575 with {\arrow[#1]{stealth}}
	}}},
	near arrow/.style={postaction={decorate,decoration={
				markings,
				mark=at position .275 with {\arrow[#1]{stealth}}
	}}},
	far arrow/.style={postaction={decorate,decoration={
				markings,
				mark=at position .800 with {\arrow[#1]{stealth}}
	}}},
}

\pgfdeclarepatternformonly{south west lines}{\pgfqpoint{-0pt}{-0pt}}{\pgfqpoint{3pt}{3pt}}{\pgfqpoint{3pt}{3pt}}{
	\pgfsetlinewidth{0.4pt}
	\pgfpathmoveto{\pgfqpoint{0pt}{0pt}}
	\pgfpathlineto{\pgfqpoint{3pt}{3pt}}
	\pgfpathmoveto{\pgfqpoint{2.8pt}{-.2pt}}
	\pgfpathlineto{\pgfqpoint{3.2pt}{.2pt}}
	\pgfpathmoveto{\pgfqpoint{-.2pt}{2.8pt}}
	\pgfpathlineto{\pgfqpoint{.2pt}{3.2pt}}
	\pgfusepath{stroke}}

\newcommand{\Tr}{\operatorname{Tr}}

\newcommand{\bbI}{\mathbb{I}}

\newcommand{\bbR}{\mathbb{R}}

\newcommand{\calT}{\mathcal{T}}

\newcommand{\eqnref}[1]{Eq.~\eqref{#1}}
\newcommand{\figref}[1]{Fig.~\ref{#1}}

\definecolor{Acolor}{RGB}{242,120,121}
\definecolor{Bcolor}{RGB}{130,230,130}
\definecolor{Ccolor}{RGB}{153,163,252}

\begin{document}

\title{From entanglement generated dynamics to the gravitational anomaly and chiral central charge}

\author{Ruihua Fan} 

\affiliation{Department of Physics, Harvard University, Cambridge, MA 02138, USA}

\begin{abstract}
    We apply modular flow---entanglement generated dynamics---to characterize quantum orders of ground state wavefunctions.
    In particular, we study the linear response of the entanglement entropy of a simply connected region with respect to modular flow.
    First, we apply it to (1+1)D conformal field theories and demonstrate its relationship to the chiral central charge---or equivalently the perturbative gravitational anomaly---which is shown to vanish. Next, we apply it to (2+1)D gapped ground states where it reduces to a recently proposed formula by Kim et. al. that is conjectured to compute the edge chiral central charge. Modular flow provides an intuitive picture for this conjecture based on bulk-edge correspondence.
    We also provide numerics on free fermion models that lend support to our picture.
\end{abstract}

\maketitle

Quantum many-body systems at zero temperature support a variety of different phases which exhibit distinct quantum orders that are encoded in the ground state wavefunction~\cite{sachdev2011quantum}.
Yet, \emph{how to extract this order from the ground state alone} is an interesting question.
For example, phases with spontaneous symmetry breaking can be detected from local order parameters. In many other cases, local observables are not sufficient to distinguish phases and quantum entanglement measures become indispensable~\cite{XiaoGangQIBook}.

One of the most widely used measures is entanglement entropy.
Specifically, the entanglement entropy of a simply connected subregion, whose linear size is larger than the correlation length if the system is gapped, encodes many universal properties of the system.
For example, its scaling with subsystem size can be used to determine the total central charge of (1+1)-dimensional critical systems~\cite{Calabrese:2004eu}, the total quantum dimension of (2+1)-dimensional gapped systems~\cite{Kitaev:2005dm,LevinWen2006}, etc. 
Nevertheless, there are several universal properties whose encoding remains to be understood.
Here, we utilize the linear response of the entanglement entropy under modular flow to generate new fingerprints of universal properties of ground state wavefunctions.
To keep the discussion general we do not assume any global symmetries, but incorporating them is relatively straightforward. 

The concept of entanglement linear response is motivated by introducing the \emph{modular Hamiltonian}~\cite{Haag:1992hx} defined as follows.
Let $\ket{\psi}$ denote the ground state wavefunction and $\rho_D = \Tr_{\bar D} \ket{\psi} \bra{\psi}$ the reduced density matrix of a subregion $D$. The modular Hamiltonian is then $K_D := -\ln \rho_D$, which is a Hermitian operator with a lowered bounded spectrum. 
Thus, the reduced density matrix $\rho_D$ can be regarded as a Gibbs state with respect to the modular Hamiltonian $K_D$ at temperature $T=1$.
With this definition, the entanglement entropy can be written as $S_D =-\Tr \rho_D \ln \rho_D = \Tr\left(K_D \rho_D \right)$, which suggests that all previous studies involving entanglement entropy can be interpreted as understanding the thermodynamic energy at a fictitious equilibrium.
The generalization that is employed here is to perturb the system away from equilibrium and examine the response.

A natural choice of perturbation is the \emph{modular flow}, the unitary evolution generated by the modular Hamiltonian
$U_D(s) := e^{-i K_D s}$, where $s\in \bbR$ is called the modular time and is dimensionless~\cite{Haag:1992hx}.
We are interested in the response of the entanglement entropy. This idea has been widely used in the study of holography ~\cite{Jafferis:2014lza,Jafferis:2015del,Chen:2018rgz} and recently discussed in condensed matter systems ~\cite{Kim:2021tse}.
Since the subregion where the modular flow is applied and that where the entanglement is measured must have an overlap to give a non-trivial answer, we denote the subregions as $AB$ and $BC$, respectively, with $B$ being the overlapping part. We have
\begin{equation}
\begin{gathered}
    \ket{\psi} \mapsto \ket{\psi(s)} \equiv e^{-i K_{AB} s} \ket{\psi} \,,\\
    S_{BC}(s) = -\Tr \left[ \rho_{BC}(s) \ln \rho_{BC}(s) \right]\,.
\end{gathered}
\end{equation}
We will focus on the linear response which can already encode non-trivial information and is defined by
\begin{equation}
\label{eq:entanglement linear response}
\frac{d}{ds} S_{BC}(s) \Big|_{s=0} = i \Tr\left( [K_{AB},K_{BC}]\rho_{ABC} \right)\,,
\end{equation}
where $K_{AB}$ and $K_{BC}$ are understood as $K_{AB} \otimes\bbI_C$ and $\bbI_A\otimes K_{BC}$.
Our analytic discussion will focus on a direct analysis of the left-hand side while the numerics will compute the right-hand side.

In this letter, we apply this idea to two related situations. One is the (1+1)-dimensional conformal field theories (CFTs), where we use it to extract the perturbative gravitational anomaly. 
This calculation immediately suggests its relevance to the (2+1)-dimensional chiral phases, the second situation that we discuss.
In this case, \eqnref{eq:entanglement linear response} reduces to the modular commutator formula, which was recently proposed to compute the chiral central charge ~\cite{Kim2021gjx}. 
We provide an argument to explain the connection between \eqnref{eq:entanglement linear response} and the chiral central charge based on the modular flow and bulk-edge correspondence. 
Numerics in both continuum and lattice free-fermion models are provided as support.

\emph{(1+1)D CFTs and gravitational anomaly:}
We consider a general (1+1)D CFT with a holomorphic and anti-holomorphic stress-energy tensor that has the following two-point functions on the complex plane
\begin{equation}
	\braket{T(z) T(0)} = \frac{c/2}{z^4}\,,\,
	\braket{\tilde T(\bar z) \tilde T(0)} = \frac{\bar c/2}{\bar{z}^4}\,,
\end{equation}
where $c$ and $\bar{c}$ are the holomorphic and anti-holomorphic central charge, their difference $c_-=c - \bar{c}$ is the chiral central charge~\cite{Ginsparg:1988ui}.
When $c_-\neq 0$, the theory has a perturbative gravitational anomaly, which means that it cannot be defined on manifolds with a boundary~\cite{ALVAREZGAUME1984269}. This condition can be derived from \eqnref{eq:entanglement linear response} as follows.

Consider a system on a circle of length $L=2\pi$. We assume that the conformal vacuum $\ket{\psi}$ can be realized as the ground state of a critical lattice system and has a Schmidt decomposition. 
Let $A,B,C$ be three adjacent disjoint intervals shown below
$$
\begin{tikzpicture}[scale=1]
\draw[gray!50,thick] (0,0) circle (1);
\draw[fill=black] (1,0) circle (0.05) node[right] {$x'=0$ \text{or} $L$};
\draw[fill=black] (-0.174,0.984) circle (0.05) node[above] {$x'=y$};
\draw[fill=black] (-0.985,-0.174) circle (0.05) node[left] {$x'=\ell$};
\draw[fill=black] (0.1736,-0.9848) circle (0.05) node[below] {$x'=x$};
\draw[<->,>=stealth,line width=1.5,Acolor!90!black] (1,0) arc (0:98:1 and 1) node at (0.9,0.9) {$A$};
\draw[<->,>=stealth,line width=1.5,Bcolor!90!black] (-0.19,0.984) arc (100:188:1 and 1) node at (-0.95,0.8) {$B$};
\draw[<->,>=stealth,line width=1.5,Ccolor!90!black] (-0.985,-0.185) arc (190:278:1 and 1) node at (-0.85,-0.95) {$C$};
\end{tikzpicture}
$$
It follows from the conformal invariance that the modular Hamiltonian of any single interval $D=[\ell_L,\ell_R]$ is a local integral of the stress-energy tensor
\begin{equation}
\label{eq:KD CFT}
	K_D = 4\pi \int_D \frac{\sin \frac{\ell_R-x}{2} \sin \frac{x-\ell_L}{\pi}}{\sin \frac{\ell_R-\ell_L}{2}}\, T_{00}(x) dx\,,
\end{equation}
where $T_{00} = \frac{1}{2\pi} (T+\tilde T)$ is the energy density ~\cite{Casini:2011kv}. Therefore, the modular flow implements conformal transformation.
The entanglement entropy $S_{BC}$ under the modular flow of $K_{AB}$ is given by the two-point function of the twist operator~\footnote{Strictly speaking, we have to multiply the twist operators by phase factors to have a real result. It is related to the nonzero momentum of the conformal ground state on the circle, which is already a signature of the anomaly. The fact that the calculation remains not self-consistent even after such a redefinition means the issue can only be moved but not hidden, which is a common feature of anomalies.}
\begin{equation}
	\lim_{n\rightarrow 1} \frac{1}{1-n} \log \braket{0|e^{iK_{AB}s} \calT_n(x) \bar\calT_n(y) e^{-iK_{AB}s}|0}
\end{equation}
where $\calT_n$ and $\bar{\calT}_n$ are the twist and anti-twist operators with the conformal dimension $h_n = \frac{c}{24}\left(n-\frac{1}{n} \right),\bar h_n = \frac{\bar c}{24}\left(n-\frac{1}{n} \right)$~\cite{Calabrese:2004eu}. 
This correlation function can be computed conveniently in the Heisenberg picture and we get
\begin{equation}
\frac{d}{ds} S_{BC}(s) \Big|_{s=0} = \frac{\pi c_-}{6} \left( 1 - 2 \frac{\sin \frac{x}{2} \sin \frac{\ell-y}{2}}{\sin \frac{\ell}{2} \sin \frac{x-y}{2}} \right)
\end{equation}
where the ratio of the four sine functions is the cross ratio of the four points depicted in the figure, and $c_-$ is the chiral central charge.
Its appearance can be understood via the quasiparticle picture, which is an exact rephrasing of the technical calculation in the current context.
The entanglement entropy of $\ket{\psi}$ comes from the chiral and anti-chiral degrees of freedom, the numbers of which are proportional to $c$ and $\bar c$, respectively.
They move across the entanglement cut at $x'=y$ in the opposite directions under the modular flow and thus give opposite contributions to the change of entanglement entropy.

One interesting case is the limit $x\rightarrow 2\pi$, at which $A,B,C$ occupy the entire circle and their union is in a pure state $\rho_{ABC}=\ket{\psi}\bra{\psi}$.
On the one hand, the result reduces to
\begin{equation}
\label{eq:1+1d cft result}
	\frac{d}{ds} S_{BC}(s) \Big|_{s=0} = \frac{\pi c_-}{6}\,.
\end{equation}
On the other hand, it follows from the Schmidt decomposition that
\begin{equation}
	K_{AB}\ket{\psi} = K_{C}\ket{\psi}\,.
\end{equation}
and the modular flow in $AB$ should not change the entanglement entropy $S_{BC}$ at all.
This implies that the chiral central charge $c_-=c-\bar{c}$ must vanish if $\ket{\psi}$ is a genuine (1+1)-dimensional wavefunction that allows for a Schmidt decomposition.
This proves the absence of the perturbative gravitational anomaly in a genuine (1+1)D system.

Our discussion only uses the ability of having a Schmidt decomposition explicitly, while the original statement of the anomaly is about the ability of having boundaries. 
Their connection can be established as follows.
Note that none of quantum field theories has a factorized Hilbert space~\cite{Witten:2018zxz}. 
In order to define the Schmidt decomposition and the reduced density matrix, one has to first embed the original Hilbert space into a larger one that does factorize. Both the larger Hilbert space and the embedding require specifying a boundary condition~\cite{Ohmori:2014eia}.
Therefore, the ability of having a Schmidt decomposition is equivalent to that of having boundaries~\cite{Hellerman:2021fla}.

A (1+1)D chiral CFT can exist on the spatial edge of a (2+1)D gapped system, where the anomaly is canceled by the bulk. In that case, it is natural to expect that applying \eqnref{eq:entanglement linear response} in a similar manner as above gives the chiral central charge of the (2+1)D system~\cite{KaneFisher1997,Cappelli:2001mp,Kitaev:2005hzj}, which is a requirement of the bulk-edge correspondence. We confirm this intuition below but the result will be shown have a small difference.

\begin{figure}[!t]
\centering
\includegraphics[width=7cm]{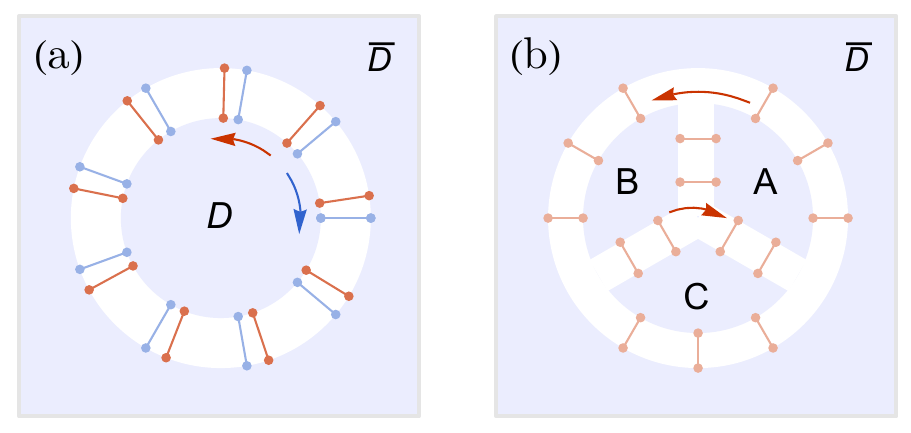}
\caption{Distribution of entangled degrees of freedom and their modular flow. (a) Red and blue bonds are the chiral and anti-chiral modes. The arrows designate their motion under the modular flow $U_D(s)$. (b) The anti-chiral degrees of freedom are suppressed for the clarity of the figure. The arrows designate the motion under the modular flow $U_{AB}(s)$.}
\label{fig:ccc ABC partition}
\end{figure}

\emph{(2+1)D gapped states and chiral central charge:}
We consider (2+1)D gapped state whose edge is described by a (1+1)D CFT with central charges $(c,\bar{c})$. The chiral central charge $c_-=c-\bar c$ is not necessarily zero. 
Then the modular Hamiltonian $K_D$ of a simply-connected subregion $D$, with a smooth boundary whose linear size is larger than the correlation length, can be approximated by a CFT Hamiltonian
\begin{equation}
\label{eq:KD of 2d bulk}
	\rho_D = \frac{1}{Z_D} e^{-K_D},\quad K_D = \beta\frac{2\pi}{\ell_D} (L_0 + \bar{L}_0 - \frac{c+\bar{c}}{24})
\end{equation}
where $L_0$ and $\bar{L}_0$ are the Virasoro generators, $\ell_D$ is the length of the boundary of $D$ and $\beta \ll 1$ is a non-universal constant that depends on the UV details~\cite{Kitaev:2005dm,HaldaneLi,Swingle:2011hu,Chandran2011,QiLudwig}.
The entanglement entropy of this region follows from the Cardy formula~\cite{Cardy:1986ie}
\begin{equation}
	S_D = \frac{\pi}{6\beta}(c+\bar c) \ell_D  + \cdots
\end{equation}
The sub-leading terms are irrelevant to our current discussion.
We interpret the above formula in the following way~\cite{QiLudwig}. The entanglement comes from chiral and anti-chiral modes sitting on the entanglement cut $\partial D$, whose line densities are given by
\begin{equation}
\label{eq:line density}
	\lambda_{\text{chiral}} = \frac{\pi}{6\beta} c,\quad
	\lambda_{\text{anti-chiral}} = \frac{\pi}{6\beta} \bar c
\end{equation}
\figref{fig:ccc ABC partition}~(a) gives a sketch, where the chiral and anti-chial modes are colored in red and blue, respectively. One caveat is that we depict them like Bell pairs just for the clarity of the figure. The actual entanglement structure is much more complicated and cannot be captured by Bell pairs, especially near the triple contact regions. Otherwise, the two modular Hamiltonians $K_{AB}$ and $K_{BC}$ commute and will give a trivial result~\footnote{We thank Michael Zaletel for pointing this out.}. 
The gapped nature of the state implies that they are entangled only with their neighbors.
The motion of these modes under the modular flow depends on their chiralities. The chiral modes that are inside $D$ will move towards the left, which is shown by the red arrow in \figref{fig:ccc ABC partition}~(a). The anti-chiral ones respond in the opposite direction, which is shown by the blue arrow \figref{fig:ccc ABC partition}~(a). For simplicity, the following discussion focuses on the chiral modes and the contribution from the anti-chiral modes is the opposite.

Now, we apply the above picture to compute the entanglement linear response in the geometry shown in \figref{fig:ccc ABC partition}~(b), where the disk $D$ is divided into three parts $A,B$ and $C$ that are not necessarily of equal sizes. 
It is a direct generalization of the geometry used for \eqnref{eq:1+1d cft result} to (2+1)D. 
The chiral modes are depicted as the red bonds in \figref{fig:ccc ABC partition}~(b), whose motion under the modular flow $U_{AB}$ follows the arrows. 
Only ones near the $ABD$ and the $ABC$ triple-contact points can affect $S_{BC}$.
Near the $ABD$ triple-contact point, the chiral modes, that used to be in $A$ and entangled with $\bar D$, will move into $B$, which leads to an increment of $S_{BC}$. Similarly, the motion of the anti-chiral modes will decrease $S_{BC}$. After summing up the contribution from the $ABD$ and $ABC$ triple-contact points, we have
\begin{equation}
\label{eq:change of SBC 2d}
	\frac{dS_{BC}}{ds} \Big|_{s=0} = 2(\lambda_{\text{chiral}} -
	\lambda_{\text{anti-chiral}}) v
\end{equation}
where $v$ is the velocity, which is non-universal and given by $\beta$ defined in \eqnref{eq:KD of 2d bulk}.
Combining \eqref{eq:line density} with \eqref{eq:change of SBC 2d}, the non-universal factors cancel out and we are left with the chiral central charge \emph{only}
\begin{equation}
\label{eq:2d argument result}
	\frac{dS_{BC}}{ds}\Big|_{s=0} = \frac{\pi c_-}{3}\,.
\end{equation}
There is a factor-of-two discrepancy between \eqnref{eq:1+1d cft result} and \eqnref{eq:2d argument result}, which comes from summing over the contribution from the two triple-contact points.
Recalling the right-hand side of \eqnref{eq:entanglement linear response}, we recognize that this is exactly the modular commutator formula for the chiral central charge that was recently proposed by Ref.~\cite{Kim2021gjx} and has been shown to pass different consistency requirements.
Our derivation gives an intuitive picture for their conjecture.

We extend this argument further by applying it to other geometries and compare the prediction with numerical results that will be shown later.

\begin{figure}[t]
\includegraphics[width=7cm]{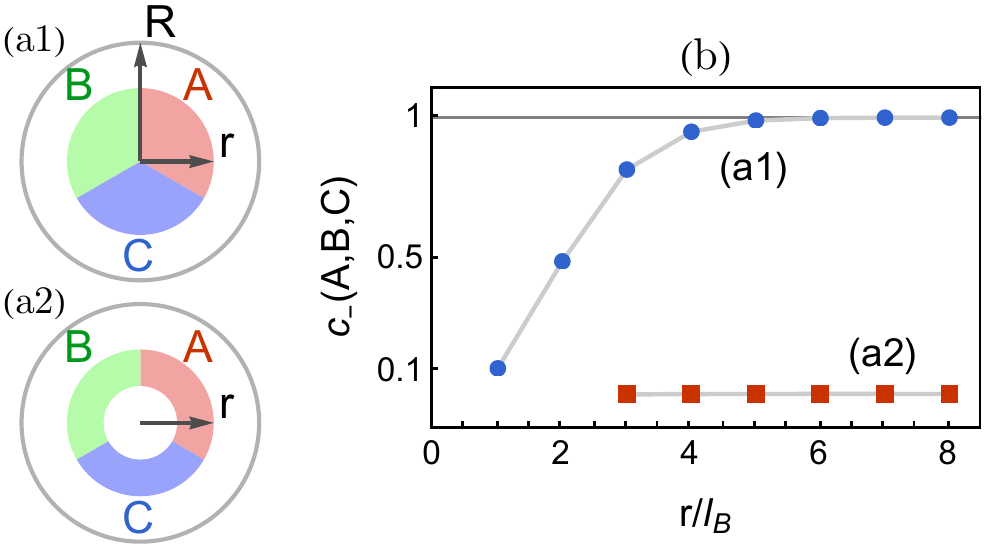}
\caption{Modular commutator of the $\nu=1$ IQH. (a) The choice of the subsystem. (b) The chiral central charge approaches $1$ when the subsystem size is large enough (but still smaller than the total system size). We choose $R=10$ and the LLL has 50 degenerate states. For the annulus geometry shown by (a2), we choose the inner radius to be $2$. }
\label{fig:cMinus IQH}
\end{figure}

When $A,B,C$ form an annulus (see \figref{fig:cMinus IQH}~(a2)), one can repeat the analysis and show that the chiral modes at the hole of the annulus now makes opposite contribution to $S_{BC}$ compared with the ones at the $ABD$ triple contact point. Therefore, we expect to have a vanishing result. This is consistent with the intuition that an annulus supports chiral modes with opposite chiralities on the inner and outer edges, the total chiral central charge of which should be zero.

\begin{figure}[b]
\includegraphics[width=6.5cm]{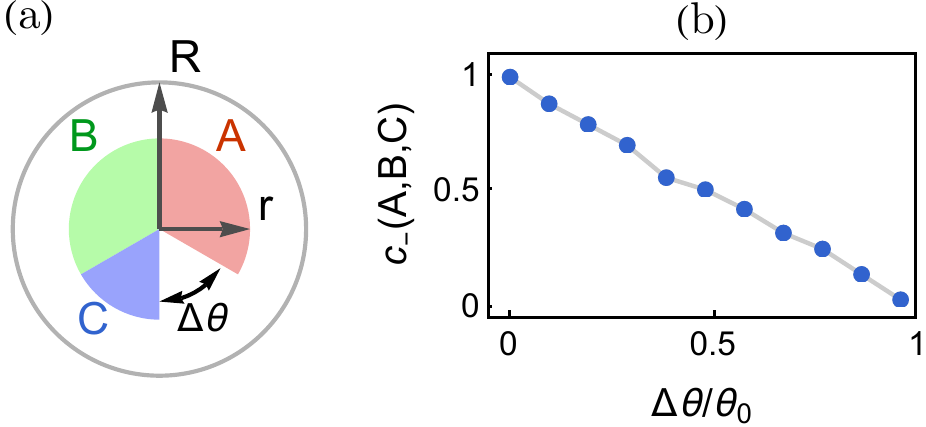}
\caption{Modular commutator of the fully-filled lowest Landau level (LLL) in a different geometry. (a) The choice of the subsystem. We choose the angle of $A$ and $B$ to be equal to $\theta_0 = 2\pi/3$. (b) The chiral central charge decreases monotonically as the angle of the subsystem $C$ decreases. }
\label{fig:IQH incomplete disk}
\end{figure}

Another interesting case is when $A,B$ and $C$ form an incomplete disk. We remove some part of the region $C$ that is adjacent to $A$, as is shown in \figref{fig:IQH incomplete disk}~(a).
Then, the region $D$ penetrates between $A$ and $C$, which gives us a quadruple-contact point.
Near this point, $D$ can have chiral modes that are entangled with region $B$. They will make opposite contribution to chiral modes entangling $B$ and $C$ and decrease the result. 
As more of the region $C$ is removed, the chiral modes between $B$ and $D$ dominates and can even cancel the contribution from chiral modes at the $ABD$ triple contact point. It eventually gives a vanishing result.

\emph{Numerics in free fermion systems:}
The numerics compute the right-hand side (the commutator) instead of the left-hand side of \eqnref{eq:entanglement linear response}. We call 
$c_-(A,B,C) = \frac{3i}{\pi}\Tr \left( \rho_{ABC}[K_{AB},K_{BC}] \right)$ 
as the numerical value of the right-hand side.

We choose the $\nu=1$ integer quantum Hall (IQH) in the continuum, which has the chiral central charge $c_-=1$.
The system is defined on a disk with a radius $R$ and a uniform magnetic field perpendicular to the plane. The ground state wavefunction is given by the Slater determinant of the lowest-Landau level wavefunctions.

We first compute $c_-(A,B,C)$ for the geometry drawn in \figref{fig:cMinus IQH}~(a1).
The result is shown by the blue dots in \figref{fig:cMinus IQH}~(b). 
The result quickly converges to and plateaus at the desired value as the radius of the disk increases and becomes much larger than the magnetic length (while still smaller than the total system size). 
One can also deform the three subsystems $A$, $B$ and $C$ in a smooth way, the result does change as long as they form a complete disk.
This provides a good evidence that \eqnref{eq:2d argument result} or the modular commutator can indeed compute the chiral central charge at the length scale larger than the correlation length (the magnetic length in this case).

We then consider the two other geometries. One is the annulus, shown in \figref{fig:cMinus IQH}~(a2). We find a null result as shown by the red squares in \figref{fig:cMinus IQH}~(b).
The other is the incomplete disk, shown in \figref{fig:IQH incomplete disk}~(a). Here we choose the angle of $A$ and $B$ to be $\theta_0 = 2\pi/3$ and decrease the angle of $C$ by $\Delta\theta\in[0,\theta_0]$.
The result decreases almost linearly from 1 to 0 as the removed angle $\Delta\theta$ increases from $0$ to $\theta_0$. 
Both results are consistent with our previous argument.

So far, our discussion is restricted to a certain phase. It is interesting to examine how $c_-(A,B,C)$ behaves across phase transitions.
This can be suitably studied with a lattice model. We choose the chiral $p$-wave superconductor~\cite{Read:1999fn,bernevig2013topological}. The Hamiltonian consists of three components: on-site chemical potential, nearest-neighbor hopping and $p$-wave superconducting pairing. Without loss of generality, we fix the hopping and pairing strength to be $t=\Delta=1$ and let the chemical potential $\mu$ be the only tunable parameter. The phase diagram has two trivial insulating phases at $\mu<0$ and $\mu>8$ and two two chiral topological phases with opposite chiralities at $0<\mu<8$.
Each chiral phase support a single chiral Majorana edge mode and thus has a chiral central charge $c_- =\pm 1/2$. 
\begin{equation*}
\begin{gathered}
	\includegraphics[width=5cm]{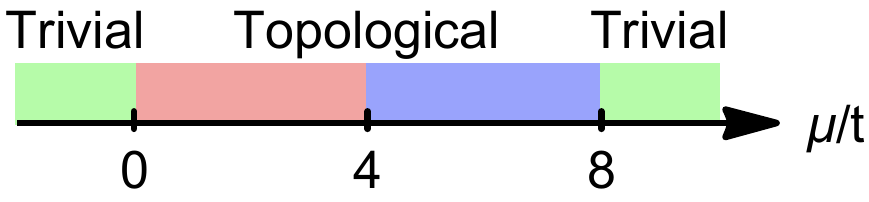}
\end{gathered}
\end{equation*}

\begin{figure}[!t]
\includegraphics[width=7cm]{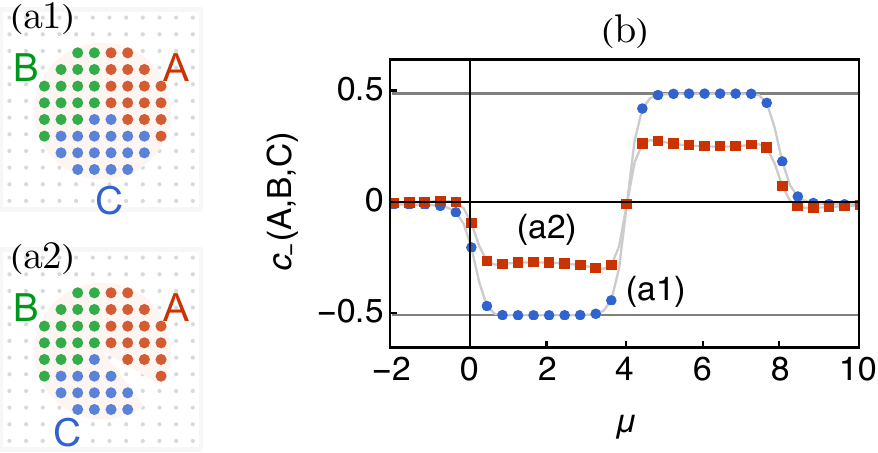}
\caption{Modular commutator of the $p+ip$ superconductor. It shows transition at $\mu=0,4,8$ respectively. (a) The choice of the subsystem. (b) The chiral central charge plateaus to $\pm1/2$ in the two topological phase when $A,B,C$ form a complete disk. We choose the system size $L_x=L_y=20$ and the radius of the disk $R=6$. For the incomplete disk, we choose the angle depletion for region $C$ to be $0.3$.}
\label{fig:cMinus SC}
\end{figure}

The result is shown in \figref{fig:cMinus SC}. For a complete disk (blue dots), it plateaus to the quantized value $\pm 1/2$ in the two topological phases and shows the multiple topological transition clearly. For an incomplete disk (red squares), the result is smaller but still detects the transition. 

\emph{Discussion:}
Our extension of the rigorous (1+1)D results to the (2+1)D scenario requires appealing to the quasiparticle picture, i.e., the change of entanglement entropy arises from the motion of quasiparticles whose number is conserved by the modular flow. 
We leave to future work a rigorous derivation of \eqnref{eq:2d argument result} from Chern Simons theories, analogous to the derivation of topological entanglement entropy. We remark that to obtain the chiral central charge from an effective theory, going beyond topological quantum field theory is needed~\footnote{we thank Xi Yin for pointing this out.}.

It will be interesting to further analyze and extend the response theory against modular flow to other examples, such as the phase transitions between (2+1)D chiral phases, situations with global symmetries or generic gapless systems. Understanding its implication on numerics will also be interesting.

\emph{Note added:} Near the completion of this work, we become aware of \cite{Zou:2022nuj}, which appears to have some overlap with the present manuscript.

\bigskip

During the time this work was in progress, I received tremendous insight and help from Rahul Sahay and Ashvin Vishwanath. Collaborating with them on a related project also helped me a lot.
I thank Ehud Altman, Yimu Bao, John Cardy, Yiming Chen, Meng Cheng, Zhehao Dai, Chunxiao Liu, Joel Moore, Bowen Shi, Yantao Wu, Huajia Wang, Xi Yin, and especially Yingfei Gu, Daniel Jafferis, Ari Turner and Michael Zaletel for many helpful discussions and comments.
I am grateful to Yingfei Gu, Rahul Sahay, and Ashvin Vishwanath for carefully reading through the manuscript and providing helpful comments.
RF is supported by a Simons Investigator award (AV) and by the Simons
Collaboration on Ultra-Quantum Matter, which is a grant from the Simons Foundation (651440, AV). RF is supported by the DARPA DRINQS program (award D18AC00033).
This work is funded in part by a
QuantEmX grant from ICAM and the Gordon and Betty Moore Foundation through Grant GBMF9616 to Ruihua Fan.

\bibliography{ref.bib}

\end{document}